\documentclass[%
 preprint,
 amsmath,amssymb,
 aps,
]{revtex4-2}

\usepackage{graphicx}
\usepackage{dcolumn}
\usepackage{bm}
\usepackage{xcolor}
\usepackage{upgreek}

\begin{document}

\title{Mid-infrared high-sensitive cavity-free in-situ CO gas sensing based on up-conversion detection}

\author{Zhao-Qi-Zhi Han$^{1,2,3}$}
\altaffiliation{These authors contributed equally to this work.}

\author{He Zhang$^{1,2,3,}$\textsuperscript{*}}
\email{zhanghe23@ustc.edu.cn}

\author{Fan Yang$^{4}$}

\author{Xiao-Hua Wang$^{1,2,3}$}

\author{Bo-Wen Liu$^{1,2,3}$}

\author{Jin-Peng Li$^{1,2,3}$}

\author{Zheng-He Zhou$^{1,2,3}$}

\author{Yin-Hai Li$^{1,2,3,5}$}

\author{Yan Li$^{1,2,3}$}

\author{Zhi-Yuan Zhou$^{1,2,3,5}$}
\email{zyzhouphy@ustc.edu.cn}

\author{Bao-Sen Shi$^{1,2,3,}$}
\email{drshi@ustc.edu.cn}

\affiliation{%
$^{1}$Laboratory of Quantum Information, University of Science and Technology of China, Hefei 230026, China\\
$^{2}$CAS Center for Excellence in Quantum Information and Quantum Physics, University of Science and Technology of China, Hefei 230026, China\\
$^{3}$Anhui Province Key Laboratory of Quantum Network, University of Science and Technology of China, Hefei 230026, China\\
$^{4}$National Key Laboratory of Electromagnetic Space Security, Tianjin 300308, China\\
$^{5}$Anhui Kunteng Quantum Technology Co. Ltd., Hefei 231115, China\\
}


\date{\today}

\begin{abstract}
Carbon monoxide (CO) is a significant indicator gas with considerable application value in atmospheric monitoring, industrial production and medical diagnosis. Its fundamental vibrational band locates around 4.6 $\upmu$m and has larger absorption line strength than that of overtone band, which is more suitable for the precise identification and concentration detection of CO. In this paper, the up-conversion detection is employed to convert the mid-infrared absorption signal obtained by TDLAS to the visible light band, then a silicon-based detector is utilized for detection. By which, we can achieve the highest sensitivity of 79.6 ppb under the condition of cavity-free in-situ with an absorption range length of only 0.14 m. Furthermore, the single-photon level real-time detection of CO concentration after the diffuse reflection is realized by using SPAD. This work demonstrates the merits of the up-conversion detection in terms of its functionality at room temperature and capacity for sensitivity detection. Furthermore, it presents a design and optimization methodology that has the potential to underpin the advancement of the method towards more practical applications, like industrial process monitoring, medical diagnosis and so on.
\end{abstract}

\maketitle

\clearpage

\section{Introduction}
CO is an important indicator gas that plays a crucial role in atmospheric pollution monitoring, industrial production and medical diagnosis. In atmospheric pollution monitoring, CO is a typical pollutant primarily derived from fossil fuel combustion and biomass burning\cite{wang2000mst, holloway2000jgr}. It not only severely impacts the ecological environment by reacting with other gases in the atmosphere and forming photochemical smog\cite{thompson1992science}, but also indirectly exacerbates global warming by affecting tropospheric ozone formation\cite{khalil1984science}. In industrial production, monitoring and analyzing CO concentration fluctuations in industrial processes allows for the optimization of workflows and key parameters, resulting in enhanced production efficiency as well as energy savings and reduced emissions\cite{wang2022ate, wei2021ate}. In medical diagnosis, CO may serve as a biomarker for the health effects of air pollution and related respiratory diseases\cite{gajdocsy2010jbr}. Besides, it is also significant for the early diagnosis and detection of cardiovascular diseases\cite{xue2025sab}. Therefore, developing efficient, real-time, and sensitive CO detection systems is of paramount importance for environmental protection and public health.

TDLAS (Tunable Diode Laser Absorption Spectroscopy) is a prominent optical method for gas concentration detection. In recent years, it has been widely adopted for measuring CO concentrations across diverse fields—such as industrial process monitoring and safety\cite{barh2019aop}, environmental and atmospheric science\cite{liu2020ac}, and medical diagnosis-owing to its ultra‑high sensitivity\cite{ghorbani2017oe}, capability for gas‑specific fingerprint identification, and ability to perform multicomponent simultaneous measurements.

The fundamental principle of TDLAS is based on the Beer-Lambert law which means the detection sensitivity of TDLAS-based sensor systems is directly proportional to the absorption line strength of the target gas. The CO molecules absorption lines strength of the second overtone band locates around 1.57 $\upmu$m, therefore a near-infrared(NIR) distributed feedback(DFB) lasers  can be employed to achieve the required sensitivity combined with multi-pass gas cells. Multi-pass cells typically have an effective optical path length of tens or even hundreds of meters, enabling to reach a detection limits at ppm or even ppb levels\cite{wagner2012apb, lin2013apb, azhar2017apb}. Notably, the absorption line strength of the fundamental vibrational band in MIR region is four orders of magnitude higher than that of the second overtone band. Therefore, MIR quantum cascade lasers(QCLs), interband cascade lasers(ICLs), or difference-frequency-generated lasers can be used in TDLAS-based sensor systems, combining with suitable photodetectors, to further enhance detection sensitivity. For TDLAS systems utilizing mid-infrared light resources, Mercury Cadmium Telluride(MCT) detectors are generally used to detect absorption intensity\cite{kosterev2002ao, tao2012oe, ren2012apb}. However, MCT detectors suffer from low detection sensitivity and strict working condition such as deep cooling, and are mostly applied in scientific research. What’s more, nonlinear frequency upconversion technology has also been proven to be an effective alternative for mid-infrared signal detection\cite{armstrong1962pr, warner1971opto}. In this technology, high-performance detectors based on wide-bandgap materials (e.g., silicon) are used to detect MIR light after frequency be converted to visible/NIR light\cite{ge2022cpb, han2024prd, huang2022nc}. 

In this paper, we demonstrated a mid-infrared highly sensitive cavity-free in situ trace CO monitoring system based on frequency upconversion and conducted comprehensive performance testing of the system, including detection limits, system response linearity, and system detection capability under single-photon sensitivity. Experimental results showed that using difference-frequency-generated mid-infrared light as the modulated source combined with an upconversion-based mid-infrared detector can greatly enhance CO detection sensitivity and accuracy. With an optical path length of 14 cm and a 70 s acquisition time, a detection limit of 79.6 ppb was achieved, while real-time detection at a sampling frequency of 1 kHz achieved a precision of 1.47 ppm. Additionally, the paper proposed a cavity-free in situ scenario requiring single-photon-level detection sensitivity. Such scenarios can include CO leak detection at arbitrary locations in factories or non-line-of-sight enclosed spaces, and real-time monitoring of exhaled CO content. This scenario was simulated by introducing diffuse reflection targets into the optical path, and the system's performance in real-time measurement of TDLAS signals at the single photon level was demonstrated experimentally.

\section{Result}
\begin{figure}
\centering\includegraphics[width=\textwidth]{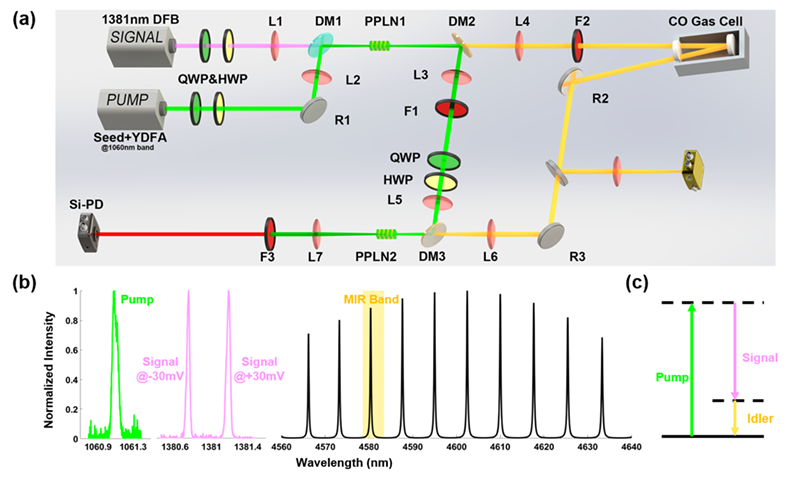}
  \caption{(a) Schematic diagram of the experimental setup. L terms: lenses; DM terms: dichromatic mirrors; F terms: filters; PPLN terms: periodically poled lithium niobate crystals; R terms: Ag mirrors; HWP: half-wave plate; QWP: quarter-wave plate; Si-PD: silicon-based photodiode detector; MCT-PD: HgCdTe photodiode detector. (b) Spectral of pump and signal under different modulation voltages, along with their corresponding MIR spectral band and CO absorption peaks. (c) Energy level diagram corresponding to the DFG process.}
  \label{fig:setup}
\end{figure}

A simple diagram of our experimental setup is displayed in \textbf{Figure \ref{fig:setup}(a)}. The generation of MIR light is achieved through the difference-frequency generation (DFG) process, while the detection of the MIR beam passing through the gas is accomplished via an upconversion detection. As illustrated in \textbf{Figure \ref{fig:setup}(b)}, the spectra of the pump laser and near-infrared laser under $-30\ mV$ and $30\ mV$ modulation, as measured by the spectrometer, are presented. The spectrometer used in experiment is predicated on its lower resolution in comparison to the spectral line width, thus only enabling the characterization of the central wavelength of the laser output. The application of the principle of energy conservation (${\lambda _{mir}} = \frac{{{\lambda _{nir}} - {\lambda _p}}}{{{\lambda _{nir}} \times {\lambda _p}}}$) enables the calculation of the wavelength coverage of the MIR beam, as illustrated by the shaded region in \textbf{Figure \ref{fig:setup}(b)}. The black line represents the theoretically calculated absorption curve for CO using the
molecular parameters tabulated in HITRAN\cite{hitran2022} and the Beer-Lambert law\cite{cope1988}. The collision and Doppler effect which can broaden line width of the molecules are taken into account when the Voigt profile are used\cite{sobel2012excitation}. It has been demonstrated that the generated MIR beam precisely covers the CO absorption of line R10 in the $1 \leftarrow 0$ band. Furthermore, it has been shown that by expanding the modulation range, it is possible to extend the MIR wavelength coverage to encompass the adjacent lines R11 and R9.

\begin{figure}
\centering\includegraphics[width=\textwidth]{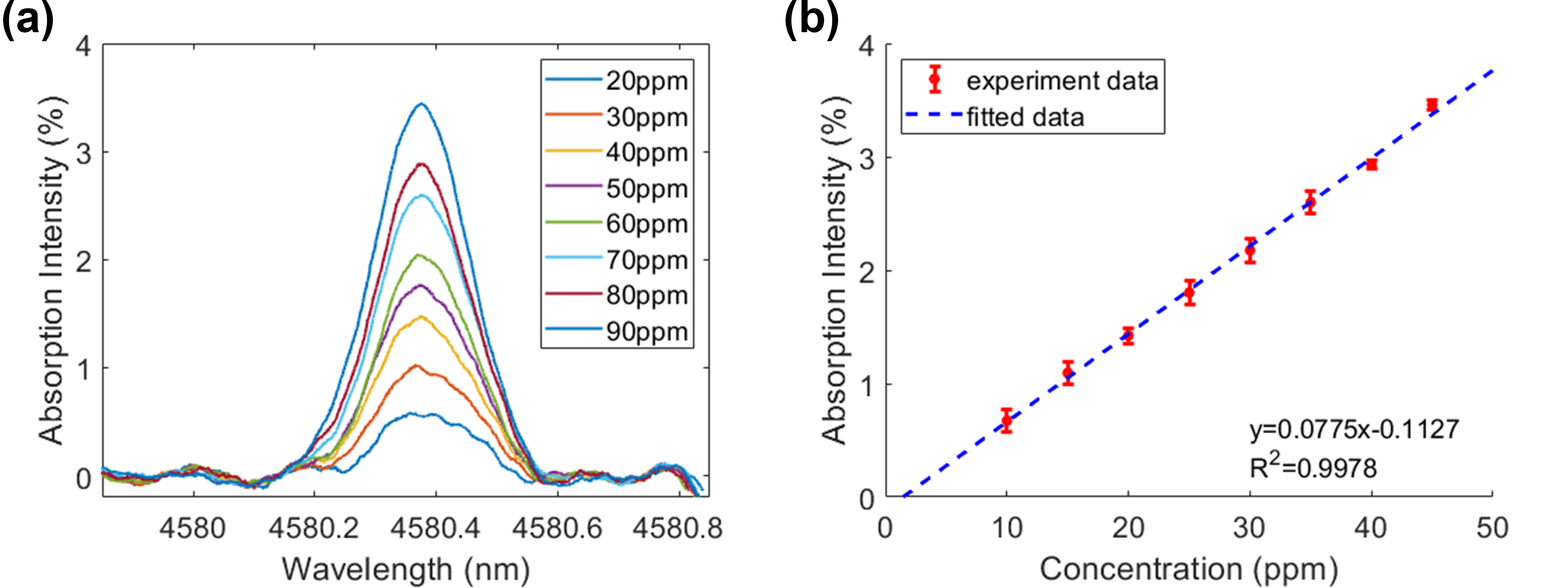}
  \caption{CO concentration measurement results using upconversion system under $\upmu$W MIR illumination. (a) Absorption line shapes corresponding to different CO concentrations. (b) Relationship between absorption rate and concentration at various CO levels, along with linear fitting results.}
  \label{fig:shape}
\end{figure}
Initially, the system is tested under $\upmu$W level MIR illumination conditions. To assess the upconversion system's response to concentration variations, nine sets of CO/He mixed gases with concentrations ranging from 10 to 45 ppm are employed. The approach employed to obtain the absorption rate results was a relatively standardized one. Specifically, an odd-degree polynomial fitting method was utilized to establish the baseline, upon which the absorption rate was calculated\cite{dong2016apl}. As illustrated in \textbf{Figure \ref{fig:shape}(a)}, the absorption profiles obtained at varying concentrations show a decline in both peak absorption intensity and total absorption area as the concentration is reduced. In order to calibrate the system's linearity to a greater extent, the average responsivity is calculated from multiple measurements. As shown by the calibration curve in \textbf{Figure \ref{fig:shape}(b)}, the system under investigation exhibits linear behaviour across the concentration range, with an absorption rate of approximately $0.0775\ \%\cdot \text{ppm}^{-1}$, a y-intercept of $0.1127\ \%$ and the R-squared for this linear fitting of 0.9978.The experimental measurement of absorption rates has been shown to be in close agreement with the results of the HITRAN simulation. For instance, the experimentally measured absorption rates at 20, 30, and 40 ppm concentration were 1.4288\ \%, 2.1752\ \%, and 2.9294\ \% respectively, corresponding to the simulated values of 1.4389\ \%, 2.1505\ \%, and 2.8571\ \%. The current modulation signal frequency is 1 kHz, and the results have been derived from processing a continuous 30 ms period, or 30-times consecutive measurements. 

\begin{figure}
\centering\includegraphics[width=100mm]{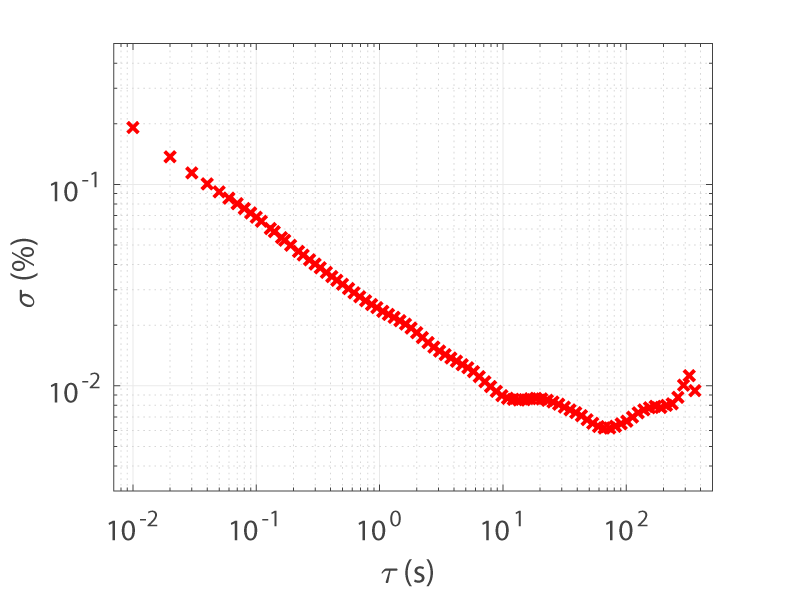}
  \caption{Allan variance of absorption rate measured by upconversion detection at 5 ppm CO.}
  \label{fig:allan}
\end{figure}

The precision and stability are investigated by means of a study into the Allan-Weller deviation, which is studied at a CO concentration of 5 ppm. This deviation is plotted against the integration time (see \textbf{Figure \ref{fig:allan}}). It shows that within a specified time frame, the accuracy of measurements can be enhanced by suppressing white noise through the augmentation of integration time. At an integration time of about 70 s, the system under consideration attains optimal measurement accuracy. This is evidenced by an absorption rate measurement precision of $6.17 \times 10^{-3}\ \%$, which corresponds to a CO concentration of 79.6 ppb$\left( {{\rm{ = }}{\raise0.7ex\hbox{${6.17 \times {{10}^{{\rm{ - }}3}}\ \% }$} \!\mathord{\left/
 {\vphantom {{6.17 \times {{10}^{{\rm{ - }}3}}\% } {0.0775\%  \cdot {\rm{ppm}}}}}\right.\kern-\nulldelimiterspace}
\!\lower0.7ex\hbox{${0.0775\ \%  \cdot {\rm{ppm}}}$}}} \right)$ calculated by the Beer-Lambert law. Furthermore, an examination of the Allan variance reveals that at 30 ms, it stands at $0.11\ \%$, a finding that corresponds with the results obtained at other concentrations, as illustrated in \textbf{Figure \ref{fig:shape}}. This indicates that the system displays favourable repeatability and stability.

\begin{figure}
\centering\includegraphics[width=100mm]{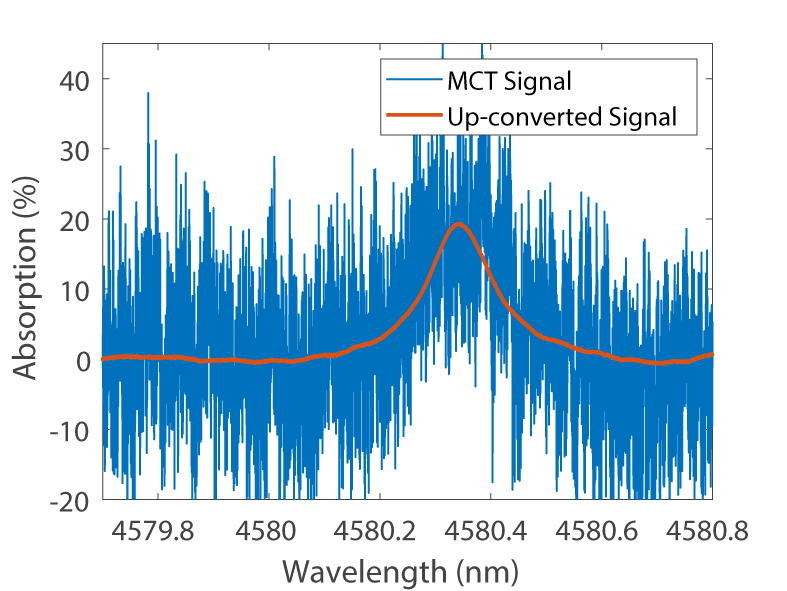}
  \caption{Comparison of measurement results between MCT and upconversion detection of 300 ppm CO at room temperature.}
  \label{fig:MCT}
\end{figure}

A comparison is made between the measurement results of the MCT (PDAVJ10, Thorlabs) and the upconversion detection under room-temperature detection conditions, as illustrated in \textbf{Figure \ref{fig:MCT}}. It was evident that under identical testing conditions (1 kHz modulation, 1 MHz sampling rate, 10 MHz low-pass filtering), the MCT-PD output signal at room temperature exhibited significant spikes. These fluctuations correspond to absorption intensity variations reaching $20\ \%$, rendering the method unsuitable for the detection of low concentrations. In contrast, the absorption curves for CO concentrations of 300 ppm, as demonstrated using the upconversion detection, reveal that this approach achieves a substantial improvement in measurement accuracy at room temperature. In comparison with the room-temperature MCT-PD, the accuracy is enhanced by two to three orders of magnitude.

\begin{figure}
\centering\includegraphics[width=\textwidth]{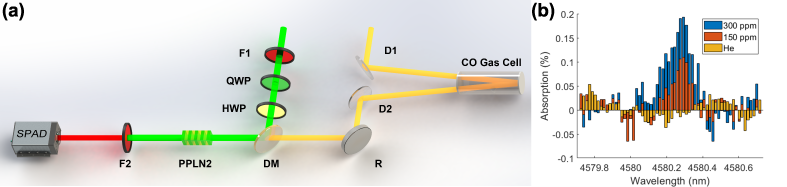}
  \caption{Single-photon-level measurement scenarios and experimental results. (a)Schematic diagram of experimental setup. D terms: diffuser. (b)Absorption line shapes of CO at different concentrations under a photon flux of approximately $1\times10^7$ cps.}
  \label{fig:concentration}
\end{figure}

\begin{figure}
\centering\includegraphics[width=\textwidth]{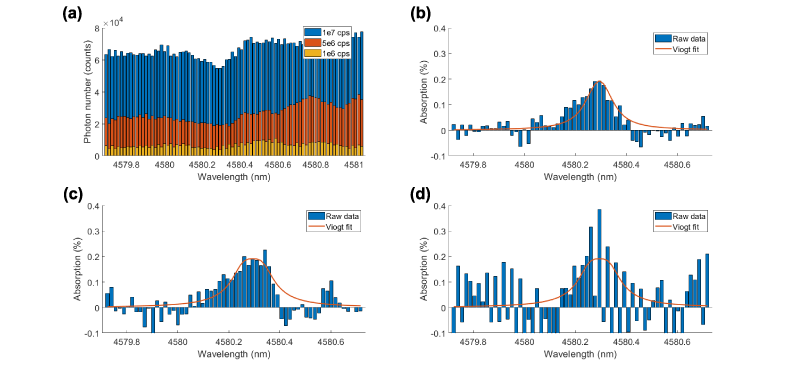}
  \caption{Absorption at 300 ppm CO. (a) Counts per bin recorded under varying photon fluxes. (b) Absorption curve at a photon flux of $1\times10^7$ cps. (c) Absorption curve at a photon flux of $5\times10^6$ cps. (b) Absorption curve at a photon flux of $1\times10^6$ cps.}
  \label{fig:counts}
\end{figure}

Furthermore, a notable benefit of the upconversion technique is its capacity to employ high-performance silicon-based detectors, thereby facilitating the detection of low-illuminance targets with high efficiency even at the single-photon level. The experiment involved the insertion of two pieces of frosted glass diffuse reflectors into the front and back of the CO cell (as shown in \textbf{Figure \ref{fig:concentration}(a)}). The MIR signal was subsequent collection by the upconversion module via diffuse reflection. The upconverted signal is then detected using a single photon avalanche diode (SPAD), simulating an in-situ gas measurement scenario in non-line-of-sight conditions. As illustrated in \textbf{Figure \ref{fig:concentration}(b)}, the absorption line shapes for varying concentrations of CO, under conditions of constant photon flux, are shown. It has been observed that the rate of absorption continues to exhibit linear variation, a finding that is consistent with the results of the simulation. The optical loss is subsequently introduced, thereby reducing the detected photon flux. As demonstrated in \textbf{Figure \ref{fig:counts}}, as the photon flux decreases, the number of photons in per bin declines, eventually leading to the obscuring of the absorption line by fluctuations. Overall, even in the single photon level, the measured concentration are in good agreement with the HITRAN reference spectrum.

\section{Discussion and Summary}
The mid-infrared spectrum encompasses a multitude of vibrational and rotational energy levels of molecules, exhibiting exceptional specificity in absorption that enables precise identification of molecules and functional groups. Furthermore, due to its remarkably high absorption intensity, it facilitates cavity-free detection at low concentrations, rendering it highly suitable for in situ monitoring applications. Additionally, upconversion detection for measuring gas concentrations offer two primary advantages. Firstly, they enable detection under entirely ambient temperature conditions, thereby eliminating the need for deep cooling. Secondly, they permit single-photon-level low-illumination detection using high-performance silicon-based detectors.

In summary, the present study investigates a highly sensitive cavity-free in-situ CO detection based on upconversion detection. In the context of micro-watt illumination, the linear relationship between the rate of absorption and the variation in concentration was examined. Allan variance measurements were used to confirm the limit of detection down to 0.0796 ppm at an absorption path length of 0.14 m with direct absorption detection. This demonstrated that upconversion techniques have the capacity to significantly enhance gas concentration measurement accuracy. Furthermore, by incorporating diffuse reflection media into the optical path, the gas was detected in non-line-of-sight conditions while attenuating illumination intensity to single-photon levels. Utilizing a SPAD, we clearly obtained the absorption outcomes under varying concentrations and photon flux, thereby confirming the system's sustained capacity for relatively precise measurements under low illumination intensity and non-cooperative target conditions. It has been demonstrated that, in general, the absorption intensity of CO gas can be accurately detected under illumination conditions ranging from micro-watt to single-photon levels. The results have been found to be in good agreement with the HITRAN simulation. This work establishes a substantial foundation for the further expansion of upconversion detection in security, environmental monitoring, and associated domains.

A fundamental parameter for evaluating gas absorption detection sensitivity is the product of absorption length and the limit of detection(LOD). The experimental findings reveal that under micro-watt illumination conditions, this metric reaches 0.011 ppm$\cdot$m, representing an enhancement of two to three orders of magnitude in comparison to the direct absorption detection results achieved with commercial room-temperature MCT-PD, and approximately fiftyfold enhancement in comparison to direct absorption detection using the same DFG light source with a deep-cooled MCT \cite{kasyutich2008apb}. In addition, the present detection capabilities of this upconversion system are comparable to those obtained by employing phase-locked amplification methods with QCLs/ICLs and MCTs under deep-cooled conditions \cite{vanderover2011apb, ghorbani2017oe}. 

There are further methods by which to enhance system performance and broaden its application scope. For instance, from the perspective of stability improvement, employing QCLs or ICLs can yield more stable and higher-power outputs, thereby further reducing the LOD. Furthermore, the benefits of DFB lasers in terms of modulation frequency and depth \cite{matsuo2018aop, silvestri2023aplp, deng2022light} can be exploited to achieve enhanced speed and an expanded dynamic range in gas concentration detection. Finally, by combining the wavelength tuning capability of the pump light, precise multi-component measurements across a broad wavelength range can be achieved (in the present system, a 1 nm shift in the 1060 nm pump wavelength corresponds to approximately 18.5 nm variation in the mid-infrared band).

\section{Method}
 The MIR beam is generated by means of the DFG process. The near-infrared signal beam is generated by a distributed feedback (DFB) laser (NLK1E5GAAA-1381nm, Wavelength Electronics) with a wavelength of approximately 1381 nm. the intense pump beam is produced by amplifying a semiconductor seed laser (KT-SL-1064, Anhui Kunteng Quantum Technology) through a Ytterbium-doped fiber amplifier (YDFA, YFA-SF-1064-50-CW, Precilasers), with a fixed center wavelength of 1061.094 nm. The PPLN1 crystal, with a length of 20 mm and a poled period of 27.21 $\upmu$m , is utilized in the DFG process. The temperature of crystal is set to be 25 $^\circ C$ within a range of $\pm$10 mK. The modulation is transferred to the MIR wavelength through the DFG process by applying modulation to the DFB laser to alter its output center wavelength.  The measurements obtained indicated that the DFB output wavelength corresponding to the center of the absorption peak is 1380.822 nm, which corresponds to a MIR center wavelength of 4580.38 nm. It is evident that both laser beams demonstrate exceptional monochromaticity, with output bandwidths measuring below 1 MHz. Consequently, it can be assumed that all lights participating throughout the detection process is monochromatic.

Following the optical filtration process, which involved the utilization of a long-pass dichroic mirror and a band-pass filter with a central wavelength of 4000 nm and a full width at half maximum (FWHM) of 2000 nm, pure MIR light is obtained. This light is then incident into the CO gas cell for single reflection. The optical path in the cell is measured at 0.14 m (the one-way length of the cell is 7 cm), the concentration of CO is calculated using the equation of state based on the air pressure (with the other components being pure He gas), and the air pressure in the cell is maintained at 10 kPa.

The MIR beam passing through CO gas is upconverted with high fidelity to approximately 861 nm in the PPLN2 crystal and detected by a Si Free-space amplified photodetector. The PPLN2 crystal is 40 mm in length with a poled period of 23 $\upmu$m. The temperature controller is set at 85.6 $^\circ C$, which has a variation range of $\pm$ 50 mK. The MIR optical power is measured to be 36.2 $\upmu$W before the DM3. The converted 861 nm beam optical power is 3.3 $\upmu$W, corresponding to a system power efficiency of ${\eta _p} = {\raise0.7ex\hbox{${{P_{vis}}}$} \!\mathord{\left/
 {\vphantom {{{P_{vis}}} {{P_{mir}}}}}\right.\kern-\nulldelimiterspace}
\!\lower0.7ex\hbox{${{P_{mir}}}$}} \times 100\%  = 9.12
\ \% $ and quantum efficiency of ${\eta _q} = \frac{{{\lambda _{vis}}}}{{{\lambda _{mir}}}}{\eta _P} = 1.72\ \% $. In order to provide additional evidence for the advantages of the upconversion detection, a MCT amplifiedphotodiode detector at ambient temperature is employed in order to detect the absorption signal (as shown within the dashed box in Figure 1). 

\begin{acknowledgments}
Funding:

We would like to acknowledge the support from National Key Research and Development Program of China (2022YFB3903102, 2022YFB3607700), National Natural Science Foundation of China (NSFC)(62435018), Innovation Program for Quantum Science and Technology (2021ZD0301100), Quantum Science and Technology-National Science and Technology Major Project (2024ZD0300800), USTC Research Funds of the Double First-Class Initiative (YD2030002023), Research Cooperation Fund of SAST, CASC (SAST2022-075) and the Opening Funding of National Key Laboratory of Electromagnetic Space Security.\\

Data availability:

Data is available from the corresponding author upon request.\\

Conflict of interest:

The authors declare no competing interests.\\

Author Contributions:

Z.-Q.-Z. Han, H. Zhang, Z.-Y. Zhou and B.-S. Shi investigated the references and came up with the idea. 
Z.-Q.-Z. Han and H. Zhang designed the experiments and were involved in building the experimental optical paths, experiment construction, data acquisition and figure visualization. 
X.-H. Wang, B.-W. Liu, J.-P. Li and Z.-H. Zhou helped collect some of the data. 
Y.-H. Li and Y. Li designed the electronics system. 
Z.-Y. Zhou and B.-S. Shi supervised the whole work and provided the funds. 
Z.-Q.-Z. Han and H. Zhang contribute equally in this work. 
All authors have contributed to writing this article. 
\end{acknowledgments}

\end{document}